\documentclass[runningheads,a4paper]{llncs}

\usepackage{amssymb}
\usepackage{graphicx}
\usepackage{caption}
\usepackage{subcaption}
\usepackage{amsmath}
\usepackage{subfig} 
\usepackage{setspace}

\newcommand{\DD}{\, \displaystyle}
\newcommand{\deff}{\, \stackrel{\text{def}}{=}}

\newcommand{\eq}[1]{\,\begin{equation}
                   #1 
                   \end{equation}
}

\newcommand{\fracc}[2]{\, \displaystyle \frac{ #1}{ #2}}

\newcommand{\morabba}[1]{\,\begin{flushright}
 \Rectsteel \\
\end{flushright}}

\newcommand{\eqq}[2]{\,\begin{equation} \label{#2}
                   #1 
                   \end{equation}
}

\newcommand{\CC}[2]{\, \binom{#1}{#2} 
}

\newcommand{\all}[2]{\,\begin{align}
                   #1 
                    \label{#2}
                   \end{align}
}

\begin{document}

\mainmatter  % start of an individual contribution

% first the title is needed
\title{ Inter-layer Degree Correlations in Heterogeneously Growing Multiplex Networks }

% a short form should be given in case it is too long for the running head
\titlerunning{ Degree Correlations in  Multiplex Networks}

% the name(s) of the author(s) follow(s) next
%
% NB: Chinese authors should write their first names(s) in front of
% their surnames. This ensures that the names appear correctly in
% the running heads and the author index.
%
\author{Babak Fotouhi$^{1,2}$, Naghmeh Momeni$^1$   }
\authorrunning{Babak Fotouhi, Naghmeh Momeni    }
% (feature abused for this document to repeat the title also on left hand pages)

% the affiliations are given next; don't give your e-mail address
% unless you accept that it will be published
\institute{$^1$ Department of Electrical and Computer Engineering\\
McGill University, Montr\'eal,  Canada\\
$^2$ Department of Sociology, 
McGill University, Montr\'eal,   Canada\\
\email{babak.fotouhi@mail.mcgill.ca} \\ 
\email{naghmeh.momenitaramsari@mail.mcgill.ca}  
 }

%\url{http://www.springer.com/lncs}

%
% NB: a more complex sample for affiliations and the mapping to the
% corresponding authors can be found in the file "llncs.dem"
% (search for the string "\mainmatter" where a contribution starts).
% "llncs.dem" accompanies the document class "llncs.cls".
%

%\toctitle{Lecture Notes in Computer Science}
%\tocauthor{Authors' Instructions}
\maketitle

\begin{abstract}
The multiplex network growth literature has been confined to homogeneous growth hitherto, where the number of links that each new incoming node establishes is the same across layers. This paper focuses on heterogeneous growth in a simple two-layer setting. 
%This paper demonstrates that non-trivial inter-layer correlations develop between   layers of a multiplex network that are independently growing, and this results holds both for preferential and uniform growth mechanisms. 
We first  analyze the case of two preferentially growing  layers and find a closed-form expression for the inter-layer degree distribution, and demonstrate that non-trivial inter-layer degree correlations  emerge  in the steady state. Then we  focus on  the case of  uniform growth. We observe that inter-layer correlations arise in the random case, too. Also, we observe that  the expression for the average layer-2 degree of nodes whose layer-1 degree is $k$, is identical for  the uniform and preferential  schemes.  Throughout, theoretical predictions  are corroborated using Monte Carlo simulations.
\end{abstract}

\section{Introduction} \label{sec:intro}
	Multiplex networks are tools for modeling networked systems in which units have heterogeneous types of interaction, making them members of distinct networks simultaneously.  The multiplex framework envisages  different layers  to model different types of  relationships between the same set of nodes. For example, we can take a sample of individuals and constitute a social media layer, in which links represent interaction on social media, a kinship layer, a geographical proximity layer, and so on. Examples of real systems that have %been conceptualized so far using the multiplex framework include citation networks~\cite{citation_aps,aps_imdb}, online social media~\cite{online}, airline networks~\cite{air}, scientific  collaboration  networks~\cite{citation_aps}, and online games~\cite{game}.
been conceptualized so far using the multiplex framework include citation networks, online social media, airline networks, scientific  collaboration  networks, and online games~\cite{survey}.

Theoretical analysis of multiplex networks  was  initiated  by the seminal papers~\cite{pion1,pion2} that invented and introduced  theoretical measures for quantifying  multiplex networks. Consequently, multiplex networks were utilized  for the theoretical study of  phenomena such as epidemics~\cite{epid1}, pathogen-awareness interplay~\cite{interplay}, percolation processes~\cite{percolation1}, evolution of cooperation~\cite{cooperation1}, diffusion processes~\cite{diffusion}  and social contagion~\cite{information}. For a thorough  review, see~\cite{survey}.

In the present paper we focus on the problem of growing multiplex networks. In~\cite{Bian1}, the   case where two layers are  homogeneously growing  (that is, the number of links that each newly-born node establishes is the same for both  layers)
according to preferential attachment is considered, and it is shown that $\overline{\ell}(k)$ (which is the average layer-2 degree of nodes whose layer-1 degree is $k$) is a function of $k$. 

%In~\cite{Bian1}, the number of links that each new incoming node establishes in both layers is equal to $m$. 

Previous results on growing multiplex networks are confined to homogeneously-growing layers~\cite{survey,Bian2,Bian1}. In the present paper, we consider heterogeneously-growing layers: each incoming node establishes $\beta_1$ links in layer 1 and $\beta_2$ links in layer 2. We also solve the problem for the case where growth is uniform, rather than  preferential. We demonstrate that, surprisingly, the expression for $\overline{\ell}(k)$ is identical to that of the preferential case. 
We verify the theoretical  findings  with Monte Carlo simulations. 

\section{Setup and Notation}

The two-layer multiplex network we consider in the present paper possesses one  set of nodes and two distinct sets of links. The network comprises two layers, corresponding to the two sets of links. Each node resides in both layers. The degree of node $x$ in layer 1 is denoted by $k_x$, and its degree in layer 2 is denoted by $\ell_x$. The number of nodes at time $t$ is denoted by $N(t)$ and the number of links at layer $i$ is denoted by $L_i(t)$, and $N_{k \ell}(t)$ is the number of nodes that have degrees $k$ and $\ell$ at time $t$. We denote the fraction of these nodes by  $n_{k,\ell}(t)$. Each incoming node establishes $\beta_1$ links in layer 1 and $\beta_2$ links in layer 2. 

At the inception, there are $L_1(0)$ links in the first layer and $L_2(0)$ links in the second layer. The network grows by the successive addition of new nodes. Each node establishes $m$ links in each layer. So the number of links in layer $i$ at time $t$ is $L_i(0)+\beta_i t$.

\section{Model 1: Preferential Attachment  }

In the first model, incoming nodes choose their destinations  according to the preferential attachment mechanism  posited in~\cite{BA}. The  probability that an existing node (call it $x$)  receives  a layer-1   link from the newly-born  node is proportional to ${k_x }$, and similarly, the probability for it to receive a layer-2 link is proportional to ${\ell_x }$.   Note that to obtain the normalized link-reception probabilities at time $t$ , the former should be divided by $L_1(0)+2\beta_1 t$ and the latter should be divided by $L_2(0)+2\beta_2 t$---the  number of links in the first and second layers, respectively.

The  addition  of a new node  at time $t$  can alter the   values of   $N_{k, \ell}$. If a node with layer-1 degree $k-1$  and layer-2 degree  $\ell$   receives a  layer-1 link,  its layer-1 degree  increments to $k$, and $N_{k \ell}$ increments as a consequence. If a node with layer-1 degree $k$   and  layer-2 degree $\ell-1$  receives a link,  its layer-2 degree   increments and  consequently,  $N_{k, \ell}$ increments. There are two events which would result in a decrease in $N_{k, \ell}$: if a node  with layer-1 degree  $k$  and layer-2 degree  $\ell$   receives a link in either  layer. Finally, each incoming node has an initial layer-1 degree and layer-2 degree of $\beta$, and increments $N_{\beta1 ,\beta2}$  when it is introduced. The following rate equation  quantifies the evolution of the expected value of $N_{k ,\ell}$  upon the introduction of a single node by addressing the aforementioned events with their corresponding probabilities of occurrence:
 \all{
  N_{k, \ell} (t+1)
 & = N_{k, \ell} (t)
+
\beta_1  \fracc{(k-1 ) N_{k-1, \ell}(t)- k   N_{k \ell}(t) }{L_1(0)+ 2\beta_1  t}
\nonumber \\
&
+
 \beta_2  \fracc{(\ell-1) N_t(k,\ell-1)- \ell  N_t(k,\ell) }{L_2(0)+ 2\beta_2 t}
 +
\delta_{k \beta_1} \delta_{\ell \beta_2}  
 .}{rate_1}

Alternatively, we can write the  rate equation for $n_{k \ell}$. Using the substitution   ${N_{k \ell}=(N(0)+t) n_{k \ell}}$, we  obtain
 \all{
& \big[ N(0)+t \big] \big[ n _{k,\ell}(t+1) - n_{k,\ell}(t) \big] 
+ n _{t+1}(k,\ell) 
 = 
 \nonumber \\
 &
+
\beta_1  \fracc{(k-1 ) N_{k-1, \ell}(t)- k   N_{k \ell}(t) }{L_1(0)+ 2\beta_1  t}
\nonumber \\
&
+
 \beta_2  \fracc{(\ell-1) N_t(k,\ell-1)- \ell  N_t(k,\ell) }{L_2(0)+ 2\beta_2 t}
 +
\delta_{k \beta_1} \delta_{\ell \beta_2}  
 .}{rate_2}

Now we focus on the  limit as ${t \rightarrow \infty}$, when the values of ${n_{k \ell}}$ reach steady states, and we  have   
\all{
\begin{cases}
\DD \lim_{t \rightarrow \infty} \beta_1 \frac{N(0)+t}{L_1(0)+ 2\beta_1 t} = \fracc{1}{2 } \\  
\DD \lim_{t \rightarrow \infty} \beta_2  \frac{N(0)+t}{L_2(0)+ 2\beta_2  t} = \fracc{1}{2} 
\end{cases}
.}{limits_1}

In this limit~\eqref{rate_2} transforms into  
\all{
n_{k \ell} = &\fracc{(k-1 ) n_{k-1,\ell} -  k n_{k \ell}}{2 } 
+ \fracc{(\ell-1 ) n_{k ,\ell-1} - \ell n_{k \ell}}{2 }  +
\delta_{k \beta_1} \delta_{\ell \beta_2} 
,}{difference_1}

Rearranging the terms, this can be equivalently expressed as follows
\all{
n_{k \ell} =  \fracc{k-1}{k+\ell+2} n_{k-1,\ell} 
\fracc{\ell-1}{k+\ell+2} n_{k, \ell-1} 
  + \fracc{2
\delta_{k \beta_1} \delta_{\ell \beta_2} }{2+\beta_1+\beta_2}
.}{difference_1}

This difference equation is solved in Appendix~\ref{app:sol_1}. The  solution  is
\all{
 n_{k,\ell}  = 
\fracc{2\beta_1(\beta_1+1) \beta_2(\beta_2+1)}{(2+\beta_1+\beta_2) k(k+1)\ell(\ell+1)}
\fracc{\CC{\beta_1+\beta_2+2}{\beta_1+1} }{\CC{k+\ell+2}{k+1}}
\CC{k-\beta_1+\ell-\beta_2}{k-\beta_1}
.
}{nkl_FIN_1}

\begin{figure}[t]
        \centering
      \begin{subfigure}[b]{.48  \textwidth}
           \includegraphics[width= 1.1 \textwidth,height=5cm]{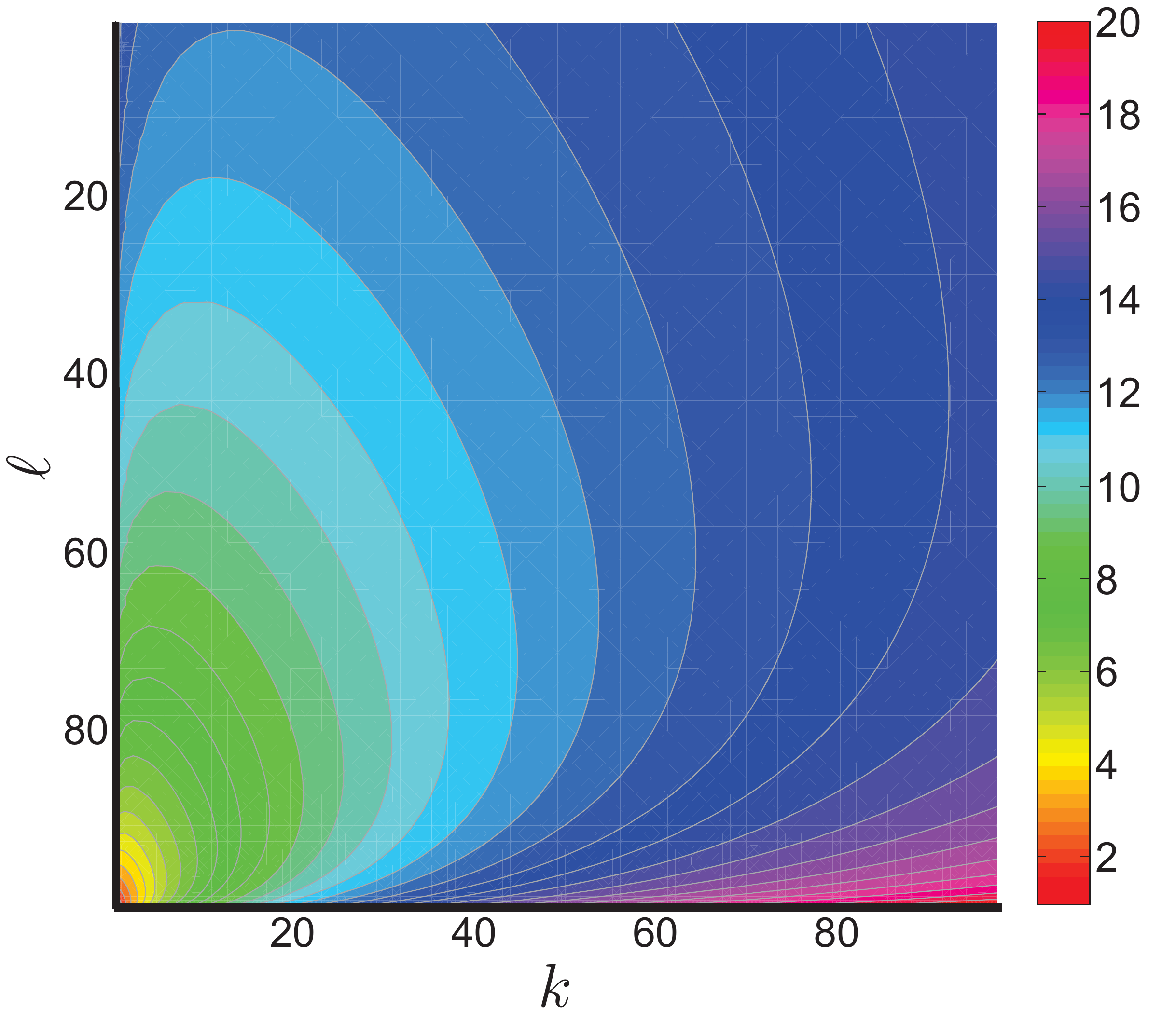}
                \caption{
The inter-layer joint degree distribution for preferential growth  with $\beta_1=2$ and $\beta_2=4$, as given by Equation~\eqref{nkl_FIN_1}. The function decays fast in $k$ and $\ell$, so we have depicted the logarithm of the inverse of this function, for better visibility. Note  the skew in the contours. Had $\beta_1$ and $\beta_2$ been equal,   the distribution would be symmetric. The function attains its maximum at $k=\beta,1$ and $\ell=\beta_2$. 
}
                \label{fig_3}
        \end{subfigure}%
           %add desired spacing between images, e. g. ~, \quad, \qquad etc.
          %(or a blank line to force the subfigure onto a new line)
   ~~~     \begin{subfigure}[b]{0.48\textwidth}
              \includegraphics[width= .99  \textwidth,height=5cm]{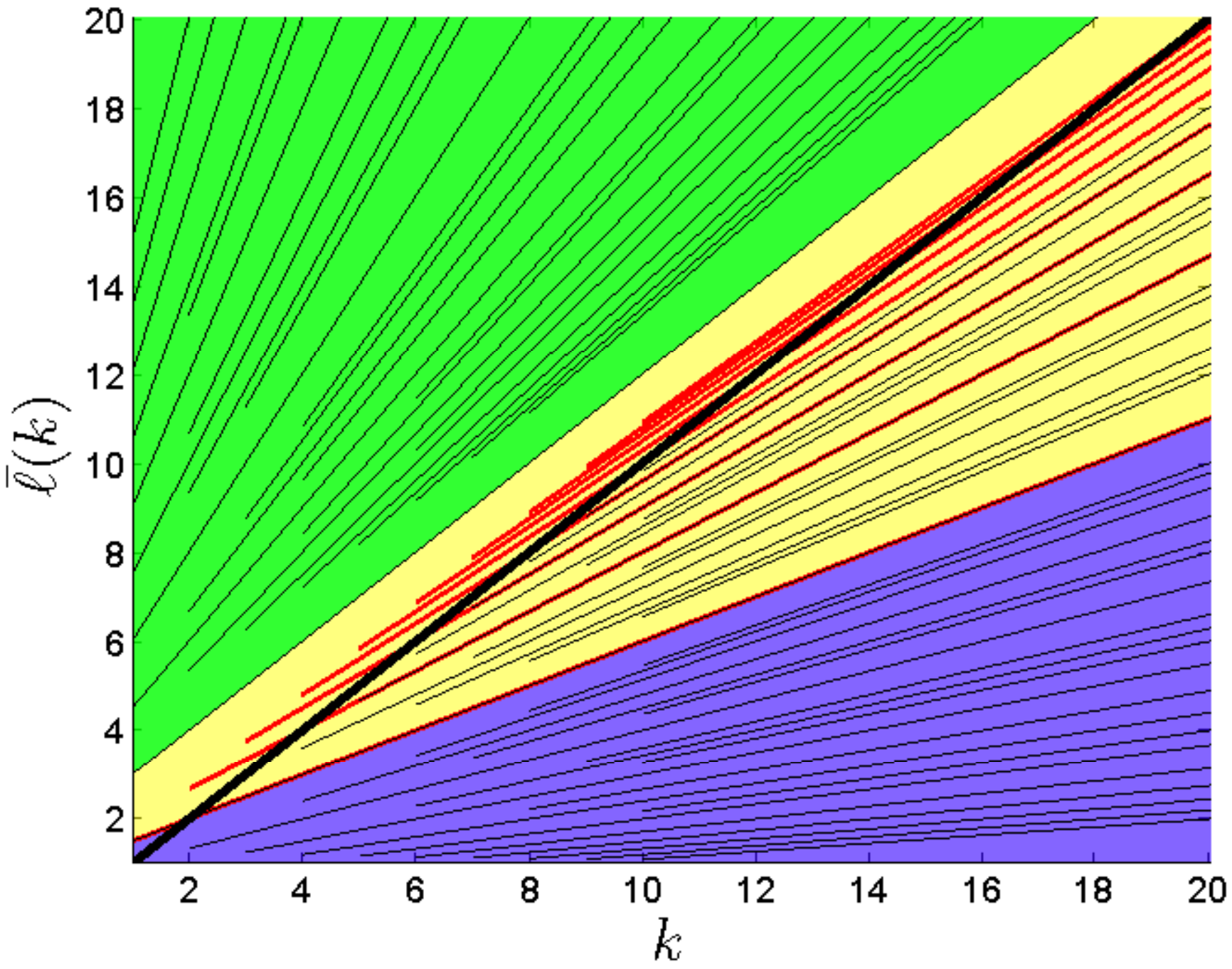}  
                \caption{ $\overline{\ell}(k)$ for  all combinations of $1\leq \beta_1,\beta_2 \leq 10$. There are three distinct regions. In the green region, $\overline{\ell}(k)>k$ regardless of $k,\beta_1,\beta_2$. In the purple region, the  converse is true. In the yellow region,  $\overline{\ell}(k)>k$ up to some critical degree $k_c(\beta_1,\beta_2)$, and above the critical degree, $\overline{\ell}(k)<k$. The top boundary corresponds to the case of $\beta_2=2,\beta_1=1$ and the bottom one pertains to $\beta_1=\beta_2=1$. 
}
                \label{fig_4}
        \end{subfigure}% 
\caption{Inter-layer joint degree distribution for preferential growth. The left figure also applies to the case of uniform growth.  symmetric. 
}
\end{figure}

This is depicted in Figure~\ref{fig_3}.
As a measure of correlation between the two layers,  we find the average layer-2 degree of the nodes whose layer-1 degree is $k$. Let us denote this quantity by $\bar{\ell}(k)$. To calculate $\bar{\ell}(k)$, we need to perform the following summation:
\all{
\bar{\ell}(k) &= \DD \sum_{\ell} \ell n_{\ell|k}= \DD \sum_{\ell} \ell \fracc{n_{k,\ell}}{n_{k}}
\nonumber \\
&=
\DD \sum_{\ell} \ell \fracc{\frac{2\beta_1(\beta_1+1) \beta_2(\beta_2+1)}{(2+\beta_1+\beta_2) k(k+1)\ell(\ell+1)}
\frac{\CC{\beta_1+\beta_2+2}{\beta_1+1} }{\CC{k+\ell+2}{k+1}}
\CC{k-\beta_1+\ell-\beta_2}{k-\beta_1}}{\frac{2\beta_1(\beta_1+1)}{k(k+1)(k+2)}}
\nonumber \\ &
= 
\DD \sum_{\ell} 
\frac{ \beta_2(\beta_2+1)(k+2) }{(2+\beta_1+\beta_2) (\ell+1) }
\frac{\CC{\beta_1+\beta_2+2}{\beta_1+1}\CC{k-\beta_1+\ell-\beta_2}{k-\beta_1 } }{\CC{k+\ell+2}{k+1}}
\nonumber \\ &
= 
\DD \sum_{\ell} 
\frac{ \beta_2(\beta_2+1) }{(2+\beta_1+\beta_2)  }
\frac{\CC{\beta_1+\beta_2+2}{\beta_1+1}\CC{k-\beta_1+\ell-\beta_2}{k-\beta_1 } }{\CC{k+\ell+2}{\ell}}
}{lbar_k_1}

In Appendix~\ref{app:nk_1}, we perform this summation. The answer is
\all{
\bar{\ell}(k)= \fracc{\beta_2}{\beta_1+1} (k+2)
.}{lbar_1}
%we prove that the following holds
%\all{
%\sum_{\ell=0}^{\infty} n_{k \ell} = \fracc{2 \beta (\beta+1)}{k (k+1)(k+2)} u(k-\beta)
%,}
%which is the degree distribution of layer 1, and is consistent with the previous results in the literature~\cite{}. 
In the special case of $\beta_1=\beta_2=m$, this reduces to $\frac{m(k+2)}{1+m}$, which is consistent with the previous result in the literature~\cite{Bian1}. 

Note that~\eqref{lbar_1} if we take the expected value of~\eqref{lbar_1}, we obtain
\all{
\DD \sum_k \overline{\ell}(k) p(k)= \fracc{\beta_2}{\beta_1+1} (\overline{k}+2)= \fracc{\beta_2}{\beta_1+1} (2\beta_1+2)=2\beta_2
,}{lbar_kol}
which  coincides with the mean degree in layer 2. 

Now let us analyze how adding a layer affects inequality in degrees. We ask, what is the probability that a node has higher degree in layer 2 than in layer 1 (on average)? That is, we seek $P(k < \overline{\ell}(k))$.   Analyzing  the inequality $k <  \frac{\beta_2}{\beta_1+1} (k+2)$, we observe that if $\beta_2<\beta_1$, then for every $k$ the inequality holds,  if $\beta_2>\beta_1$, then $k$ must be less  than $k_c=\frac{2\beta_2}{\beta_1+1-\beta_2}$. So a node with degree below $k_c$ is on average more connected in layer 2 than in layer 1. Note that since the minimum degree in layer 1 is $\beta_1$, we should impose an additional constraint on $k_c$, namely, $k_c \geq \beta_1$. This leads to $\beta_2 \leq  \beta_1-\frac{\beta_1}{\beta_1+2}$. Since   $\beta_1$ and $\beta_2$ can only take integer values, since yields $\beta_2<\beta_1$. So in order for a node with degree $k$ to have greater expected degree in layer 2 than its given degree in layer 1,  first we should have $\beta_2<\beta_1$, and second, $k\leq k_c$. In short, there are three distinct cases to discern: \textbf{(a) }If $\beta_2>\beta_1$, the inequality holds for all $k$, that is, on average, every node is more connected in layer 2 than in layer 1. \textbf{(b) } If $\beta_2 < \beta_1$, then the inequality never holds. That is, everyone is on average more connected in layer 1. \textbf{(c)} If $\beta_1=\beta_2=m$, then for nodes whose degree in layer 1 is smaller than $2m$ (which coincides with $\overline{k}$), the inequality holds, and for others it does not. So in the case of homogeneous growth, nodes whose degree in one layer is below the mean degree are on average more connected in the other layer, and nodes with degree higher $2m$ are on average less connected in the other layer. These three cases are depicted in Figure~\ref{fig_4}. The purple area pertains to case (a), where curves are $\overline{\ell}(k)$ are always  below  $k$, regardless of $\beta_1$ and $\beta_2$. The green area corresponds to case (c), where $k$ is always above $\overline{\ell}(k)$. The middle region is the one that $\overline{\ell}(k)$ curves for the cases of $\beta_1=\beta_2=m$ reside in. Those curves are depicted in red. It is visible that for each red curve, there is a cutoff degree above which $\overline{\ell}(k)<k$.

%%%%%%
%%%%%%\begin{figure}[t] 
%%%%%%\centering
%%%%%%              \includegraphics[width=.6  \textwidth,height=6cm]{fig_3}
%%%%%%%                \label{fig_3}
%%%%%%%        \end{subfigure}%
%%%%%%%           %add desired spacing between images, e. g. ~, \quad, \qquad etc.
%%%%%%%          %(or a blank line to force the subfigure onto a new line)
%%%%%%%     ~~~   \begin{subfigure}[b]{0.48\textwidth}
%%%%%%%                \includegraphics[width=  \textwidth,height=5cm]{fig_3}
%%%%%%%                \caption{
%%%%%%% 5
%%%%%%%}
%%%%%%%                \label{fig_4}
%%%%%%%        \end{subfigure}% 
%%%%%%\caption{$\overline{\ell}(k)$ for  all combinations of $1\leq \beta_1,\beta_2 \leq 10$. There are three distinct regions. In the green region, $\overline{\ell}(k)>k$ regardless of $k,\beta_1,\beta_2$. In the purple region, the  converse is true. In the yellow region,  $\overline{\ell}(k)>k$ up to some critical degree $k_c(\beta_1,\beta_2)$, and above the critical degree, $\overline{\ell}(k)<k$. The top boundary corresponds to the case of $\beta_2=2,\beta_1=1$ and the bottom one pertains to $\beta_1=\beta_2=1$.
%%%%%%}\label{fig_3}
%%%%%%\end{figure}

\section{Model 2: Uniform Attachment in both Layers}

In this model, we assume that each incoming node establishes  links  in both layers by selecting destinations from existing nodes uniformly at random. 
The rate equation~\eqref{rate_2} should be modified   to the following:
 \all{
& \big[ N(0)+t \big] \big[ n _{k,\ell}(t+1) - n_{k,\ell}(t) \big] 
+ n _{t+1}(k,\theta,\ell) 
 = 
 \nonumber \\
 &
+
\beta_1  \fracc{ N_{k-1, \ell}(t)-   N_{k \ell}(t) }{N(0)+    t}
+
 \beta_2  \fracc{  N_t(k,\theta,\ell-1)- N_t(k,\theta,\ell) }{N(0)+  t}
 +
\delta_{k \beta_1} \delta_{\ell \beta_2}  
 .}{rate_2_u}

Using the substitution $n_{k , \ell}(t) = \frac{N_{k \ell}(t)}{N(0)+t}$, this becomes
\all{
& \big[ N(0)+t \big] \big[ n _{k,\ell}(t+1) - n_{k,\ell}(t) \big] 
+ n _{t+1}(k,\theta,\ell) 
 = 
\nonumber \\ &
\beta_1  \fracc{ N_{k-1, \ell}(t)-   N_{k \ell}(t) }{N(0)+    t}
+
 \beta_2  \fracc{  N_t(k,\theta,\ell-1)- N_t(k,\theta,\ell) }{N(0)+   t}
 +
\delta_{k \beta_1} \delta_{\ell \beta_2}  
 .}{rate_3_u}

 In the steady state, that is, in  the limit as ${t \rightarrow \infty}$, this becomes
\all{
n_{k \ell}=\beta_1 \fracc{n_{k-1,\ell}-n_{k,\ell}}{1}+\beta_2\fracc{n_{k,\ell-1}-n_{k,\ell}}{1}+  \delta_{k,\beta_1} \delta_{\ell,\beta_2}.
}{difference_2_temp}
This can be simplified and equivalently expressed as follows
\all{
n_{k,\ell}= \fracc{\beta_1}{1+\beta_1+\beta_2}  n_{k-1,\ell} + \fracc{\beta_2}{1+\beta_1+\beta_2} n_{k,\ell-1}
 + \fracc{   \delta_{k,\beta_1} \delta_{\ell,\beta_2}}{1+\beta_1+\beta_2}.
}{difference_2}
This difference equation is solved in Appendix~\ref{app:sol_2}. The solution is
\all{
n_{k, \ell} = \fracc{  \beta_1^{k-\beta_1} \beta_2^{\ell-\beta_2}  \CC{k-\beta_1+\ell-\beta_2}{k-\beta_1}}{(1+\beta_1+\beta_2)^{k-\beta_1+\ell-\beta_2+1}}
}{nkl_FIN_2}

To find the conditional average degree,  that is, $\bar{\ell}(k)$, we first need the degree distribution of single layers in order to constitute the conditional degree distribution. This is found previously for example in~\cite{Bian1,ME_PRE}. The degree distribution in the first layer is  
$   n_k= \frac{1}{\beta_1 } \left( \frac{\beta_1}{\beta_1+1} \right)^{k-\beta_1+1}   $.
We need to compute
\all{
\bar{\ell}(k) &= \DD \sum_{\ell} \ell n_{\ell|k}= \DD \sum_{\ell} \ell \fracc{n_{k,\ell}}{n_{k}}
 =
\DD \sum_{\ell} \ell \fracc{ \fracc{  \beta_1^{k-\beta_1} \beta_2^{\ell-\beta_2}  \CC{k-\beta_1+\ell-\beta_2}{k-\beta_1}}{(1+\beta_1+\beta_2)^{k-\beta_1+\ell-\beta_2+1}}}{\frac{1}{\beta_1 } \left( \frac{\beta_1}{\beta_1+1} \right)^{k-\beta_1+1}}
\nonumber \\ &
= 
\fracc{  (\beta_1+1)^{k-\beta_1+1}}{(\beta_1+\beta_2+1)^{k-\beta_1+1}}
 \DD \sum_{\ell} \ell
\frac{ \beta_2^{\ell-\beta_2} \CC{k-\beta_1+\ell-\beta_2}{k-\beta_1 } }{(1+\beta_1+\beta_2)^{ \ell-\beta_2}}
}{lbar_2}

We have performed this summation in Appendix~\ref{app:nk_2}. 
The result is
\all{
\bar{\ell}(k) =
\frac{\beta_2}{\beta_1+1}(k+2)
.}{lbark_uu}

This is identical to~\eqref{lbar_1}.

\section{Simulations}
\begin{figure}[t]
        \centering
      \begin{subfigure}[b]{.48  \textwidth}
              \includegraphics[width=  \textwidth,height=5cm]{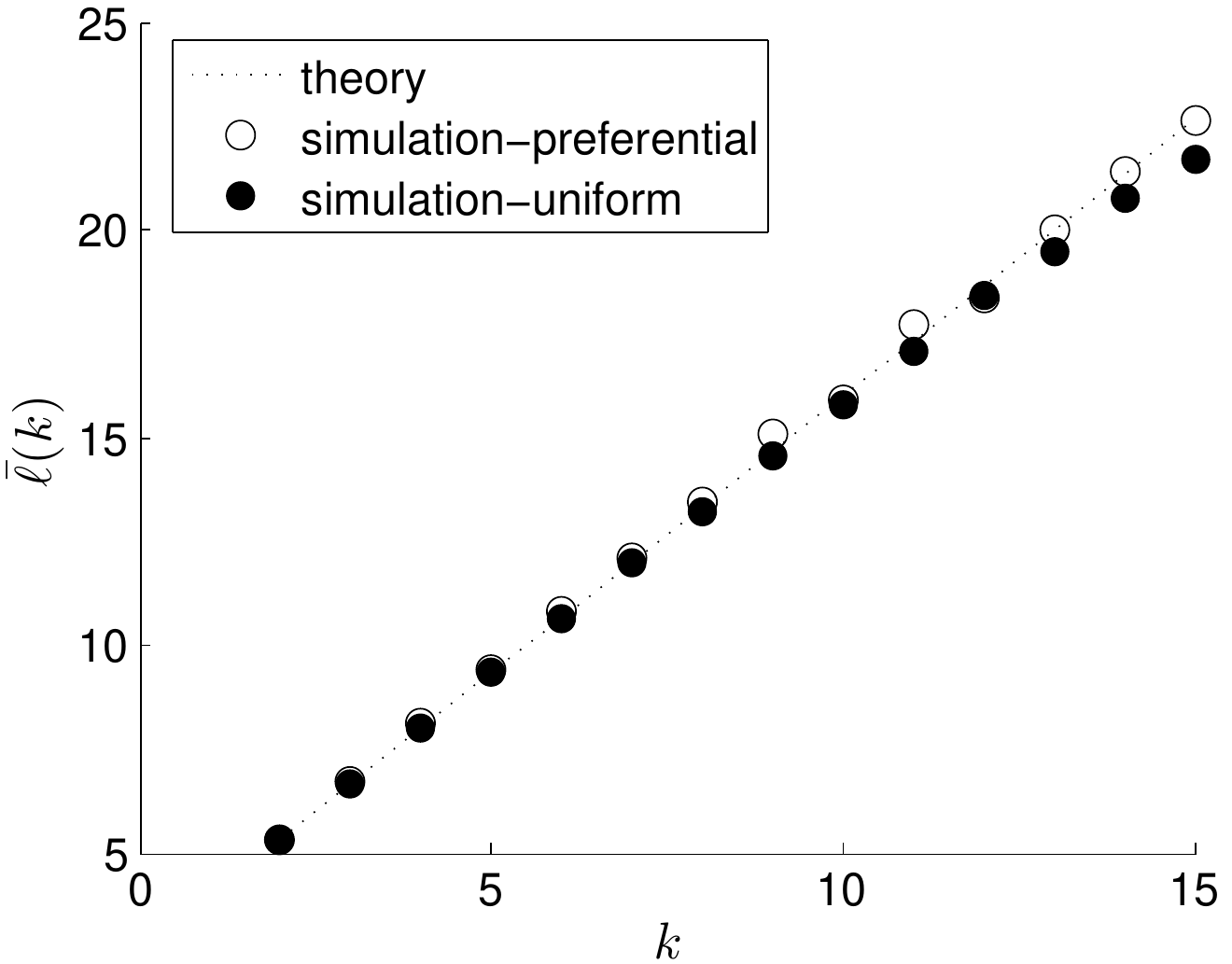}
                \caption{
  $\beta_1=2,\beta_2=4$. 
}
                \label{fig_1}
        \end{subfigure}%
           %add desired spacing between images, e. g. ~, \quad, \qquad etc.
          %(or a blank line to force the subfigure onto a new line)
     ~~~   \begin{subfigure}[b]{0.48\textwidth}
                \includegraphics[width=  \textwidth,height=5cm]{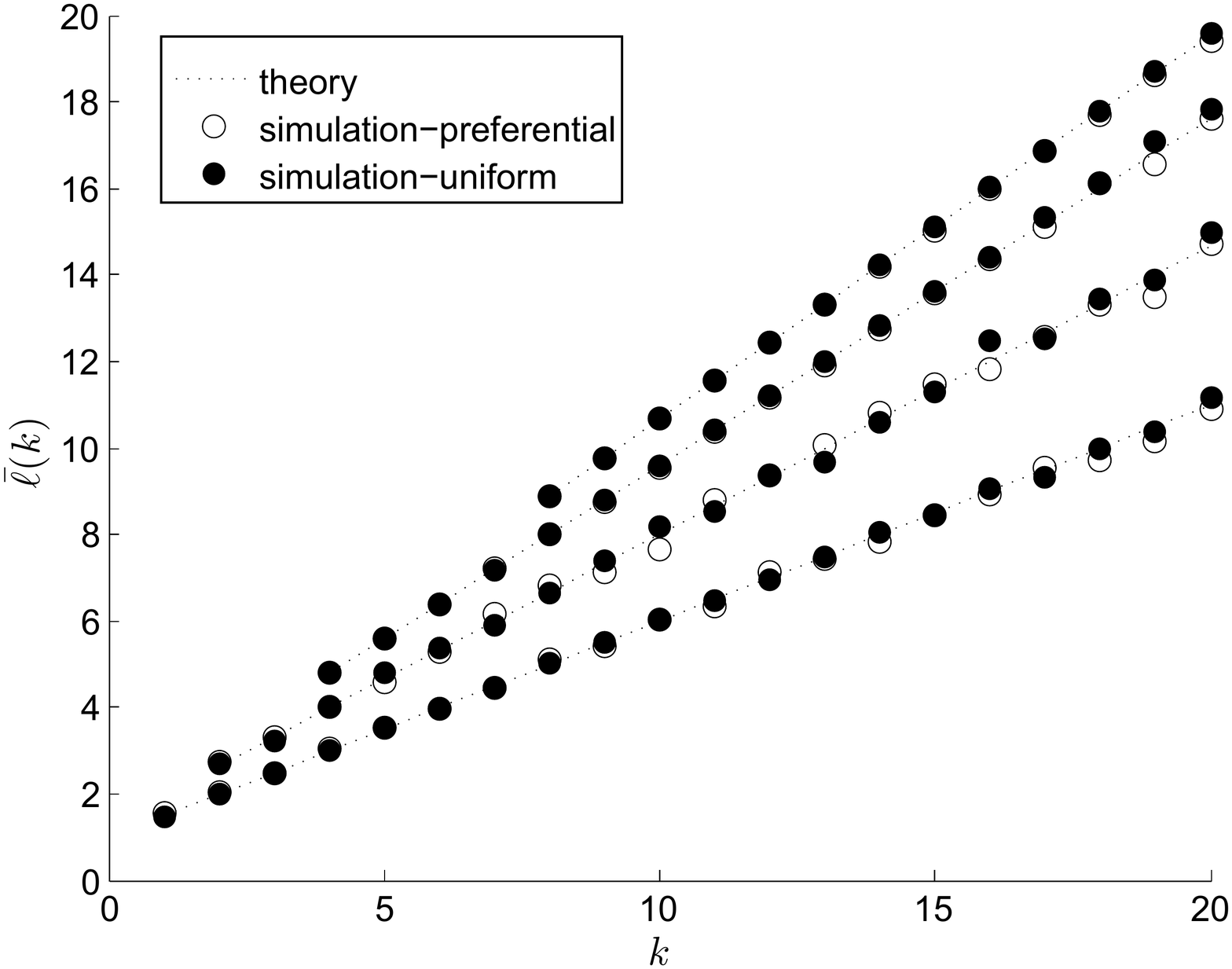}
                \caption{
$\beta_1=\beta_2=m$, for $m=1,2,4,8$. 
}
                \label{fig_2}
        \end{subfigure}% 
\caption{$\overline{\ell}(k)$ for preferential and uniform growth.  The left figure depicts $\overline{\ell}(k)$ for an example configuration of heterogeneous growth (i.e., $\beta_1 \neq \beta_2$). The right figure  represents results for homogeneous growth. It depicts different $\overline{\ell}(k)$ curves obtained for different values of $m$, where $\beta_1=\beta_2=m$ (the top line is for $m=8$, and the bottom-most line is for $m=1$). It can be seen that the slope of  $\overline{\ell}(k)$ increases as $m$ increases. 
The results are averaged over 500 Monte Carlo Trials.
}
\end{figure}

We performed Monte Carlo simulations to verify the results. Figure~\ref{fig_1} depicts $\bar{\ell}(k)$ as a function of $k$ for both uniform and preferential attachment for $\beta_1=2,\beta_2=4$. The two curves are visibly linear and overlapping. Figure~\ref{fig_2} depicts $\bar{\ell}(k)$ for both uniform and preferential attachment for $\beta_1=\beta_2=m$ for the cases $m=1,2,4,8$. It can be observed from Figure~\ref{fig_2}  that in all cases the curves for preferential and uniform growth overlap, and that the slope increases  as $m$ increases. This is consistent with the predictions of~\eqref{lbark_uu} and~\eqref{lbar_1}, where the slope is given by $\frac{m}{m+1}$. This attains its minimum at $m=1$, and reaches unity for $m \rightarrow \infty$.

\section{Summary and Future Work}
We studied the problem of multiplex network growth, where two layers were heterogeneously growing. We considered the cases of preferential and uniform growth separately. We obtained the   inter-layer joint degree distribution for both settings. We calculated $\overline{\ell}(k)$, and observed that it is identical in both scenarios. We corroborated the theoretical findings with Monte Carlo simulations. 

While the average degree $\overline{\ell}(k)$ are calculated to be the same in Eqs. (8) and (16), it does not mean the two cases have entirely the same correlation properties. Note, for example, that it was obtained in~\cite{kim}  that the two cases have different inter-degree correlation coefficients.

Plausible extensions of the present analysis are as follows. First, there is no closed-form solution in the literature for the inter-layer joint degree distribution of growing multiplex networks with nonzero coupling, where the link reception probabilities in one layer depends on the degrees  in both layers. Second, it would be informative to analyze the growth problem in arbitrary times, to grasp the finite size effects and  to  understand   how $\overline{\ell}(k)$ evolves over time, and how the time evolution differs in the preferential and uniform settings. Third, it would plausible to endow the nodes with initial attractiveness, that is, to consider a shifted-linear kernel for the preferential growth  mechanism. Fourth, a more realistic and practical model would require intrinsic fitness values for nodes, so it would be plausible to analyze the multiplex growth problem with intrinsic fitness. Finally, since most real systems are multi-layer, it would be plausible to extend the bi-layer results to arbitrary $M>2$ layers.

\begin{spacing}{.9}

\end{spacing}

\appendix

\vspace{-5mm}
\section{Solving  Difference Equation~\eqref{difference_1}}\label{app:sol_1}

We need to solve
\all{
n_{k \ell} =  \fracc{k-1}{k+\ell+2} n_{k-1,\ell} 
\fracc{\ell-1}{k+\ell+2} n_{k, \ell-1} 
  + 
\fracc{2\delta_{k \beta_1} \delta_{\ell \beta_2} }{2+\beta_1+\beta_2}
.}{difference_1_app}

We define the new sequence
\all{
m_{k \ell} \deff \fracc{(k+\ell+2)!}{(k-1)!(\ell-1)!} n_{k \ell}
.}{m_1_def}

The following holds
\eq{
\begin{cases}
 \fracc{k-1}{k+\ell+2} n_{k-1,\ell}  = \fracc{(k-1)!(\ell-1)!} n_{k \ell}{(k+\ell+2)!}  m_{k-1,\ell} \\ \\
 \fracc{\ell-1}{k+\ell+2} n_{k, \ell-1}  = \fracc{(k-1)!(\ell-1)!} n_{k \ell}{(k+\ell+2)!}  m_{k,\ell-1}.
\end{cases}
}
Plugging these into~\eqref{difference_1_app}, we can recast it as
\all{
m_{k \ell} = m_{k-1,\ell}+m_{k,\ell-1}+2 \fracc{(\beta_1+\beta_2+1)!}{(\beta_1-1)!(\beta_2-1)!}\delta_{k \beta_1} \delta_{\ell \beta_2}.
}{m_diff_1}

Now define the Z-transform of sequence $m_{k,\ell}$ as follows: 
\eqq{
\begin{cases}
\psi(z,y) \deff \DD \sum_k \sum_{\ell} m_{k,\ell} z^{-k} y^{-\ell} \\  
m_{k,\ell} = \fracc{1}{(2 \pi i)^2} \DD \oint \oint \psi(z,y) z^{k-1} y^{\ell-1} dz dy
.
\end{cases}
}{psi_def}

Taking the Z transform of every term in~\eqref{m_diff_1}, we arrive at
\all{
\psi(z,y)=&
z^{-1} \psi(z,y)+y^{-1}\psi(z,y)
 +2 \fracc{(\beta_1+\beta_2+1)!}{(\beta_1-1)!(\beta_2-1)!} z^{-\beta_1} y^{-\beta_2}
.}{psi_eq_1}
This can be rearranged and rewritten as follows
\all{
\psi(z,y) = \fracc{2}{1-z^{-1}-y^{-1}} 
  \fracc{(\beta_1+\beta_2+1)!}{(\beta_1-1)!(\beta_2-1)!} z^{-\beta_1} y^{-\beta_2}}{psi_eq_2}

The inverse transform is given by
\all{
&m_{k,\ell} = 
  \fracc{2(\beta_1+\beta_2+1)!}{(\beta_1-1)!(\beta_2-1)! } \oint \oint \fracc{z^{k-\beta_1-1}y^{\ell-\beta_2-1} dz dy}{(-4\pi^2)(1-z^{-1}-y^{-1})} 
  \nonumber \\
  & 
  = 
    \fracc{2(\beta_1+\beta_2+1)!}{(\beta_1-1)!(\beta_2-1)!} \oint \oint \fracc{z^{k-\beta_1}y^{\ell-\beta_2} dz dy}{(-4\pi^2)(zy-z-y)} 
     \nonumber \\
  & 
  = 
    \fracc{2(\beta_1+\beta_2+1)!}{(\beta_1-1)!(\beta_2-1)!} \oint \oint \fracc{z^{k-\beta_1}y^{\ell-\beta_2} dz dy}{(-4\pi^2)(y-1) \big[ z-\frac{y}{y-1} \big] } 
  .}{m_1_1}
First we integrate over $z$. We get
\all{
m_{k,\ell} &= 
    \fracc{2(\beta_1+\beta_2+1)!}{(\beta_1-1)!(\beta_2-1)!}   \oint \fracc{\left(\frac{y}{y-1}\right)^{k-\beta_1}y^{\ell-\beta_2}   dy}{(2\pi i)(y-1)  } 
\nonumber \\ &
=   \fracc{2(\beta_1+\beta_2+1)!}{(\beta_1-1)!(\beta_2-1)!}   \oint \fracc{ y^{k-\beta_1+\ell-\beta_2}   dy}{(2\pi i)(y-1)^{k-\beta_1+1}  } 
  .}{m_1_2}
Now note that the residue of $\fracc{f(y)}{(y-1)^n}$ for  positive integer equals $\fracc{f^{(n-1)}(1)}{(n-1)!}$, where the numerator denotes the ${(n-1)}$th derivative of the function $f(y)$, evaluated at ${y=1}$. Also, note that the $m$-th derivative of the function $y^n$, for integer $n$ and $m$, equals $\fracc{m!}{(n-m)!}y^{n-m}$. Combining these two facts, we obtain
\all{
m_{k,\ell}  = 
       \fracc{2(\beta_1+\beta_2+1)!}{(\beta_1-1)!(\beta_2-1)!} \CC{k-\beta_1+\ell-\beta_2}{k-\beta_1}
  .}{m_1_3}
Using~\eqref{m_1_def}, we arrive at
\all{
n_{k,\ell}  = 
       \fracc{2(\beta_1+\beta_2+1)!}{(\beta_1-1)!(\beta_2-1)!}
\fracc{1}{k(k+1)\ell(\ell+1)}
\fracc{ \CC{k-\beta_1+\ell-\beta_2}{k-\beta_1}}{\CC{k+\ell+2}{k+1}}
  .}{m_1_app_FIN}

This can be equivalently expressed as follows: 
\all{
  n_{k,\ell}  = 
\fracc{2\beta_1(\beta_1+1) \beta_2(\beta_2+1)}{(\beta_1+\beta_2+2) k(k+1)\ell(\ell+1)}
\fracc{\CC{\beta_1+\beta_2+2}{\beta+1} }{\CC{k+\ell+2}{k+1}}
\CC{k-\beta_1+\ell-\beta_2}{k-\beta_1}.
}{nkl_FIN_1_app}

\section{Performing the Summation in~\eqref{lbar_k_1}}\label{app:nk_1}

We need to calculate
\all{
\bar{\ell}(k)  = 
\DD \sum_{\ell} 
\frac{ \beta_2(\beta_2+1) }{(2+\beta_1+\beta_2)  }
\frac{\CC{\beta_1+\beta_2+2}{\beta+1}\CC{k-\beta_1+\ell-\beta_2}{k-\beta_1 } }{\CC{k+\ell+2}{\ell}}.
}{lbar_k_1_app}
We use the following identity:
$
\frac{1}{\CC{n}{m}}= (n+1)   \int_0^1  t^n (1-t)^{n-m}   dt,
$
to rewrite the binomial  reciprocal of the coefficient  as follows
\vspace{-5mm}
\all{
\fracc{1}{\CC{k+\ell+2}{\ell}} = (k+\ell+3) \DD \int_0^1 t^{\ell} (1-t)^{k+2}
dt
.}{recip_binom}
Also, from Taylor expansion, it is elementary to show that
\eqq{
S_1(x,n) \deff \DD \sum_m x^m \CC{m}{n} = \fracc{x^{n}}{(1-x)^{n+1}}.
}{taylor_0}
%We can show that $S_1(x,n) \deff   \sum_m x^m \CC{m}{n} = \fracc{x^{n}}{(1-x)^{n+1}}$ using Taylor expansions.
This identity will be used in the steps below. 
Plugging~\eqref{recip_binom}  into~\eqref{lbar_k_1_app}, we have
%\begin{widetext}
\all{
\bar{\ell}(k) & = 
\DD \sum_{\ell} 
\frac{ \beta_2(\beta_2+1) }{(2+\beta_1+\beta_2)  }
\frac{\CC{\beta_1+\beta_2+2}{\beta+1}\CC{k-\beta_1+\ell-\beta_2}{k-\beta_1 } }{\CC{k+\ell+2}{\ell}}
\nonumber \\ &
\resizebox{.95\linewidth}{!}{$
= 
\fracc{ \beta_2(\beta_2+1) }{(2+\beta_1+\beta_2)  } \CC{\beta_1+\beta_2+2}{\beta_1+1}
\DD \sum_{\ell} (k+\ell+3)  \CC{k-\beta_1+\ell-\beta_2}{k-\beta_1 }  \DD \int_0^1 t^{\ell} (1-t)^{k+2}         dt
$}
\nonumber \\&
\resizebox{.95\linewidth}{!}{$
= 
\fracc{ \beta_2(\beta_2+1) }{(2+\beta_1+\beta_2)  } \CC{\beta_1+\beta_2+2}{\beta_1+1}
\DD \int_0^1 (1-t)^{k+2} t^{-k-2}  \DD \sum_{\ell} (k+\ell+3) t^{k+\ell+2}   \CC{k-\beta_1+\ell-\beta_2}{k-\beta_1 } dt
$}
\nonumber \\&
\resizebox{.95\linewidth}{!}{$
= 
\fracc{ \beta_2(\beta_2+1) }{(2+\beta_1+\beta_2)  } \CC{\beta_1+\beta_2+2}{\beta_1+1}
\DD \int_0^1 (1-t)^{k+2} t^{-k-2} 
\fracc{d}{dt} \left[ \DD \sum_{\ell}  t^{k+\ell+3}   \CC{k-\beta_1+\ell-\beta_2}{k-\beta_1 } \right] dt
$}
\nonumber \\&
\resizebox{.95\linewidth}{!}{$
= 
\fracc{ \beta_2(\beta_2+1) }{(2+\beta_1+\beta_2)  } \CC{\beta_1+\beta_2+2}{\beta_1+1}
\DD \int_0^1 (1-t)^{k+2} t^{-k-2} 
\fracc{d}{dt} \left[t^{3+\beta_1+\beta_2}  \DD \sum_{\ell}  t^{k-\beta_1+\ell-\beta_2}   \CC{k-\beta_1+\ell-\beta_2}{k-\beta_1 }  \right] dt
$}
.}{tempak}
Using~\eqref{taylor_0}, this becomes:
\all{
\overline{\ell}(k)&
\resizebox{.95\linewidth}{!}{$
=
\fracc{ \beta_2(\beta_2+1) }{(2+\beta_1+\beta_2)  } \CC{\beta_1+\beta_2+2}{\beta_1+1}
\DD \int_0^1 (1-t)^{k+2} t^{-k-2} 
\fracc{d}{dt} \left[t^{3+\beta_1+\beta_2}   \fracc{t^{k-\beta_1}}{(1-t)^{k-\beta_1+1}}  \right] dt
$}
\nonumber \\&
=
\fracc{ \beta_2(\beta_2+1) }{(2+\beta_1+\beta_2)  } \CC{\beta_1+\beta_2+2}{\beta_1+1}
\DD \int_0^1 (1-t)^{k+2} t^{-k-2} 
\fracc{d}{dt} \left[   \fracc{t^{k+\beta_2+3}}{(1-t)^{k-\beta_1+1}}  \right] dt
\nonumber \\&
\resizebox{.9 \linewidth}{!}{$
=
\fracc{ \beta_2(\beta_2+1) }{(2+\beta_1+\beta_2)  } \CC{\beta_1+\beta_2+2}{\beta_1+1}
\DD \int_0^1 (1-t)^{\beta_1} t^{\beta_2} 
  \left[ k+\beta_2+3-(1+\beta_1+\beta_2) t  \right] dt
  $}
  \nonumber \\&
  \resizebox{.95\linewidth}{!}{$
=
\fracc{ \beta_2(\beta_2+1) }{(2+\beta_1+\beta_2)  } \CC{\beta_1+\beta_2+2}{\beta_1+1}
  \left[ (k+\beta_2+3) \DD \int_0^1 (1-t)^{\beta_1} t^{\beta_2}  dt 
-(1+\beta_1+\beta_2)  \DD \int_0^1  (1-t)^{\beta_1} t^{\beta_2+1}  dt 
  \right]  
  $}
 \nonumber \\&
 \resizebox{.95\linewidth}{!}{$
\stackrel{\textnormal{\eqref{recip_binom}}}{=}
\fracc{ \beta_2(\beta_2+1) }{(2+\beta_1+\beta_2)  } \CC{\beta_1+\beta_2+2}{\beta_1+1}
  \left[ (k+\beta_2+3)  \fracc{\beta_1! \beta_2! }{(\beta_1+\beta_2+1)!}
-(1+\beta_1+\beta_2)   \fracc{\beta_1!( \beta_1+1)! }{(\beta_1+\beta_2+2)!}
  \right]  
  $}
 \nonumber \\&
=
\frac{ \beta_2(\beta_2+1) \beta_1 ! \beta_2 !}{(2+\beta_1+\beta_2)  (1+\beta_1+\beta_2)!} \CC{\beta_1+\beta_2+2}{\beta_1+1}
  \left[ (k+\beta_2+3)  
-(\beta_2+1)
  \right]  
 \nonumber \\&
=
\fracc{\beta_2}{\beta_1+1} (k+2)
}{lbar_k_1_app}
%\end{widetext}

\section{Solving Difference Equation~\eqref{difference_2}}\label{app:sol_2}

Let us repeat the equation we need to solve for easy reference
\all{
n_{k,\ell}= \fracc{\beta_1}{1+\beta_1+\beta_2}  n_{k-1,\ell} + \fracc{\beta_2}{1+\beta_1+\beta_2} n_{k,\ell-1}
 + \fracc{   \delta_{k,\beta_1} \delta_{\ell,\beta_2} }{1+\beta_1+\beta_2}
.}{difference_2_app}

Let us define the following quantities from brevity:
%\eqq{
%\begin{cases}
%q_1 &\deff  \fracc{\beta_1}{1+\beta_1+\beta_2} \\ 
%q_2 &\deff \fracc{\beta_2}{1+\beta_1+\beta_2}
%\end{cases}
%}{qs_def}
\all{
q_1  \deff  \fracc{\beta_1}{1+\beta_1+\beta_2}~~~~,~~~~q_2  \deff \fracc{\beta_2}{1+\beta_1+\beta_2}
}{qs_def}

Taking the Z transform from both sides of~\eqref{difference_2_app}, we get
\all{
\psi(z,y)= q_1 z^{-1} \psi(z,y) + q_2 y^{-1} \psi(z,y) + \fracc{  z^{-\beta_1} y^{-\beta_2}}{1+\beta_1+\beta_2}
.}{psi2_temp_1}
This can be rearranged and recast as
\all{
\psi(z,y) = \fracc{1}{1-q_1 z^{-1} - q_2 y^{-1}} \fracc{z^{-\beta_1} y^{-\beta_2}}{1+\beta_1+\beta_2}
.}{psi2_temp_2}

This can be inverted through the following steps
\all{
n_{k \ell} &=
  \fracc{1}{(1+\beta_1+\beta_2)(2 \pi i)^2}  \DD \oint \psi(z,y) z^{k-1} y^{\ell-1} dz dy
 \nonumber \\ &
 =  \fracc{1}{(1+\beta_1+\beta_2)(2 \pi i)^2} 
\DD \oint \oint   \fracc{  z^{k-\beta-1} y^{\ell-\beta-1}}{1-q_1 z^{-1} - q_2 y^{-1}} dz dy
\nonumber \\ &
=  \fracc{1}{(1+\beta_1+\beta_2)(2 \pi i)^2} 
\DD \oint \oint  \fracc{  z^{k-\beta_1} y^{\ell-\beta_2}}{z y - y q_1 - z q_2} dz dy
\nonumber \\ &
=  \fracc{1}{(1+\beta_1+\beta_2)(2 \pi i)^2} 
\DD \oint \oint  \fracc{  z^{k-\beta_1} y^{\ell-\beta_2}}{z  -  \frac{y q_1}{y-q_2} } \fracc{1}{y-q_2} dz dy
.}{tempak_2}
There is a single simple  pole   at $z=\frac{y q_1}{y-q_2}$,  which renders the integral trivial:
\all{
n_{k \ell} &=
  \fracc{ \oint  \frac{   y^{\ell-\beta_2}}{y-q_2} \left(  \frac{y q_1}{y-q_2} \right)^{k-\beta_1} dz dy}{(1+\beta_1+\beta_2)(2 \pi i) } 
=  \fracc{  q_1^{k-\beta_1} \DD \oint  \fracc{   y^{k-\beta_1+\ell-\beta_2}}{(y-q_2)^{k-\beta_1+1}}  dz dy }{(1+\beta_1+\beta_2)(2 \pi i) } 
\nonumber \\ &
\resizebox{.98\linewidth}{!}{$
= \fracc{  q_1^{k-\beta_1} (k-\beta_1+\ell-\beta_2)!}{(1+\beta_1+\beta_2)(k-\beta_1)! (\ell-\beta_2)!}
  q_2^{\ell-\beta_2}
=\fracc{  q_1^{k-\beta_1} q_2^{\ell-\beta_2}}{(1+\beta_1+\beta_2)} \CC{k-\beta_1+\ell-\beta_2}{k-\beta_1}
 .$}
}{nkl2_app}
After inserting the expressions for $q_1,q_2$ from~\eqref{qs_def}, this becomes
\all{
n_{k,\ell}= \fracc{  \beta^{k-\beta_1} \beta_2^{\ell-\beta_2}  \CC{k-\beta_1+\ell-\beta_2}{k-\beta_1}}{(1+\beta_1+\beta_2)^{k-\beta_1+\ell-\beta_2+1}}.
}{nkl2_app_FIN}

\section{Performing the Summation in~\eqref{lbar_2}}\label{app:nk_2}
We need to perform the following summation:
\all{
\bar{\ell}(k) &= 
\fracc{  (\beta_1+1)^{k-\beta_1+1}}{(\beta_1+\beta_2+1)^{k-\beta_1+1}}
 \DD \sum_{\ell}  \ell
\frac{ \beta_2^{\ell-\beta_2} \CC{k-\beta_1+\ell-\beta_2}{k-\beta_1 } }{(1+\beta_1+\beta_2)^{ \ell-\beta_2}}
}{lbar_2_app}

Let us denote $k-\beta_1$ by $k'$ and $\ell-\beta_2$ by $\ell'$.  Also let us denote $\frac{\beta_2}{1+\beta_1+\beta_2}$ by $x$. We need to evaluate  the following sum: ${\sum_{\ell'} (\ell'+\beta_2) x^{\ell'} \CC{k'+\ell'}{k'}}$. Let us use~\eqref{taylor_0} and define  $ 
S_1(x,n) \deff   \sum_m x^m \CC{m}{n} = \frac{x^{n}}{(1-x)^{n+1}}
$. 
We have: 
\all{
&\sum_{\ell'}(\beta_2  +  \ell' )  x^{\ell'} \CC{k'+\ell'}{k'}
=  \beta_2 x^{-k'} S_1(x,k') + x  \sum_{\ell'}\ell'   x^{\ell'-1} \CC{k'+\ell'}{k'}  
\nonumber \\ &
\resizebox{.9\linewidth}{!}{$
=  \beta_2 x^{-k'} S_1(x,k') + x \fracc{d}{dx} \Big(   x^{-k'} S_1(x,k') \Big) 
=  \beta_2 x^{-k'} \fracc{x^{k'}}{(1-x)^{k'+1}}  + x \fracc{d}{dx} \Big(   \fracc{x^{k'}}{(1-x)^{k'+1}}\Big) 
$}
\nonumber \\ &
= \fracc{1}{(1-x)^{k'+2}} \big[\beta_2+ x(k'+1-\beta_2) \big] 
.}{sum11}

Replacing $x$ with $\frac{\beta_2}{1+\beta_1+\beta_2}$   and inserting this result into~\eqref{lbar_2_app}, we get
\all{&
\fracc{1}{[1-(\frac{\beta_2}{1+\beta_1+\beta_2})]^{k-\beta_1+2}}
 \big[\beta_2+\frac{\beta_2}{1+\beta_1+\beta_2}  (k-\beta_1+1-\beta_2) \big] 
 \nonumber \\ 
 & 
 = \fracc{  (1+\beta_1+\beta_2)^{k-\beta_1+2} }{(1+\beta_1)^{k-\beta_1+2}}
 \big[\beta_2+2+\frac{\beta_2}{1+\beta_1+\beta_2}  (k-\beta_1+1-\beta_2) \big] 
 \nonumber \\ 
&
 = \fracc{  (1+\beta_1+\beta_2)^{k-\beta_1+2} }{(1+\beta_1)^{k-\beta_1+2}}
 \big[\frac{\beta_2(k+2)}{1+\beta_1+\beta_2} \big] 
}{sum_11_2}
Plugging this into~\eqref{lbar_2_app}, we get

\all{
\bar{\ell}(k) &= 
\frac{\beta_2(k+2)}{1+\beta_1}
}{lbar2_app_FIN}

\end{document}